\documentclass{desyproc}
\usepackage{subfigure}
\begin{document}
\title{Solar X-rays from Axions: Rest-Mass Dependent Signatures}

\author{{\slshape Konstantin Zioutas$^{1}$, Mary Tsagri$^{1,2}$, Yannis Semertzidis$^{3}$, Thomas Papaevangelou$^4$, Antonios Gardikiotis$^1$, Theopisti Dafni$^5$\footnote{E-mail address:Theopisti.Dafni@cern.ch} \; and Vassilis Anastassopoulos$^1$}\\[1ex]
$^1$University of Patras, Patras, Greece\\
$^2$CERN, 1211 Geneve 23, Switzerland \\
$^3$Brookhaven National Laboratory, Upton, NY, USA\\
$^4$ IRFU, Centre d' \'{E}tudes Nucl´eaires de Saclay, Gif-sur-Yvette, France\\
$^5$ Laboratorio de F\'isica Nuclear y Astropart\'iculas, Universidad de Zaragoza, Zaragoza, Spain}


\desyproc{DESY-PROC-2009-05}
\acronym{Patras 2009} 
\doi  

\maketitle

\begin{abstract}
The spectral shape of solar X-rays is a power law. The more active the Sun is, the less steep the distribution. This behaviour can be explained by axion regeneration to X-rays occurring $\sim$400\,km deep into the photosphere. Their down-comptonization reproduces the measured spectral shape, pointing at axions with rest mass m$_{\rm a}\sim$17\,meV/c$^{2}$, without contradicting astrophysical-laboratory limits. Directly measured soft X-ray spectra from the extremely quiet Sun during 2009 (SphinX mission), though hitherto overlooked, fit the axion scenario.

\end{abstract}

\section{General considerations}
A first, rough comparison between the visible Sun ($\sim$5800\,K) and the well studied infant Universe \cite{Zioutas:2009bw} at a rather similar temperature ($\sim$3000\,K) is interesting due to the striking contrast between the perfect blackbody distribution and the equivalent one from the Sun. If the predicted and measured tiniest fluctuations of the cosmic plasma of $\Delta$T/T$\sim$10$^{-5}$ provide(d) fundamental new physics, one is even more tempted to conclude that the unpredictable and huge solar atmospheric fluctuations ($\Delta$T/T$\sim$10$^{3}$), of otherwise unknown origin, might well be the imprints of hidden new physics beyond the standard (solar) model. One fundamental difference to be noted is the quasi-zero magnetic field in the cosmic plasma versus the unpredictably varying huge-sized solar magnetic fields in the Tesla scale. Is this already an indication for axions or the like? We follow this simplified but observationally driven question, arriving at atypical axion signatures in solar soft and hard X-rays. Remarkably, solar X-ray emission, above its steady component, follows spatio-temporally magnetic activity. Though, the (quiet) Sun X-rays are conservatively unexpected from a cold star as our Sun, and this is the solar coronal heating problem (since 1939), which otherwise remains {\it `one of the most perplexing and unsolved problems in astrophysics to date'} \cite{Zioutas:2009bw}. To put it differently, there is no lack of problems with solar X-ray emission. This and this kind of reasoning was behind the motivation of our approach of the solar axions or other particles with similar properties.
\section{Solar magnetism as proxy for the CAST magnet}
The CAST configuration is expected to transform one day solar axions to X-rays. Such processes can take place also elsewhere, e.g. at the Sun. Thus, the working principle of the short man-made axion helioscopes might be at work in large scales near the solar surface, thanks to the ubiquitous solar magnetism. These places can act then as natural axion-to-photon converters, which for some occasionally (spatiotemporally) occurring parameter `fine tuning' might be much more efficient than any earth bound axion experiment. In fact, the rapid radial density change of the static Sun plus its wide dynamical excursions might well perform an accordingly wide plasma density scan that includes the axion resonance, i.e. the density which matches the axion rest mass ($\hbar \omega_{\rm pl}\approx m_{\rm a}c^2$). In addition, near the solar surface, this can give rise to a rather large coherence length ($>$1$-$10\,km) due to the large photon mean free path length associated with the low density and low Z solar gas. Inside a specific solar layer, the resonance condition can indeed restore coherence over large distances, provided the axion rest mass is above $\sim$10\,meV/c$^{2}$. Both, a large oscillation length and a large transverse magnetic field component maximize the axion to photon conversion ($\propto$B$^2$L$^2$). And, it is this solar synergism which earth axion experiments try actually to restore, though with inherent practical limitations.
\section{Solar axion(-like) signatures}
None analog solar X-ray spectrum is even remotely similar to that expected from converted solar axions as they are outstreaming from the hot core \cite{Zioutas:2009bw}. Instead, all analog solar spectra, be it from the quiet Sun, be it from its (non-)flaring active regions, follow an actually `colorless' power law shape, at least at first sight. But, is this expectation correct? To check this, we performed Monte Carlo simulation with the Geant4 code, which allows to follow the propagation of magnetically converted axions to X-rays below the solar surface. We argued in \cite{Zioutas:2009bw} how puzzling solar behaviour still fits the axion scenario, revising the so far widely mentioned picture \cite{carlson}, which predicts a bright X-ray spot, at the solar disc centre, from coherently converted pseudoscalars outstreaming from the solar core. But such a spot has not been detected as yet. We arrive, however, at different conclusions after comparing our simulation results with solar X-ray observations, which originate from the whole magnetic solar surface, pointing at the photosphere and/or the (lower) chromosphere as the actual axion-to-photon conversion layer. Thus, following the simulation of the propagation of X-rays in the outer layers of the Sun \cite{Zioutas:2009bw}, the observed solar spectral shape can result from an axion conversion occurring in the upper solar sub-surface; the emerging X-rays undergo a down-comptonisation while propagating in a random walk towards the 'visible' solar surface, i.e. before escaping in free space isotropically. The concluded depth of the conversion was at about 350\,km underneath the surface implying a rest mass of the axions or axion-like particles of $\sim$17\,meV/c$^{2}$. Note, in contrast, previous work \cite{carlson} assumed the pseudoscalar conversion to occur above the solar surface, implying a rest mass (far) below $\sim$10$^{-4}$\,eV/c$^{2}$. Thus, there is no contradiction between these two otherwise complementary approaches, though their spectral shape and spatial origin of the X-rays from the Sun are completely different. In the past, the various solar X-ray activities have been overlooked as being axion in origin, because of their 'wrong' spectral shape and topology, while their strong brightness could not fit our quasi-prejudice that a signature from a dark matter particle candidate should be extremely faint.

Here we give a few more atypical solar axion signatures (for more details see section 6.2 in \cite{Zioutas:2009bw}). In fact, the $\sim$10\,MK hot solar corona above non-flaring active regions is a remarkably high temperature and of potential interest. It is worth mentioning that the corresponding quiet Sun corona temperature is `only' 1-2\,MK and  that of the dramatic flares is about 10-20\,MK, i.e. it is not much different, while neither the source nor acceleration mechanisms of the particles involved have been understood. Then, it is not unreasonable to assume that in all locations, i.e. quiet Sun, flaring and not-flaring active regions, a similar mechanism might be at work. Thus, also the otherwise unexpected soft X-ray emission from the quiet Sun can be driven by the same axion-regeneration mechanism, but occurring a little deeper into the solar subsurface. This explains, then, why the shape of the quiet Sun analog spectrum is similar, though much steeper and feebler due to more (in)elastic interactions \cite{Zioutas:2009bw}. Interestingly, helioseismology unravels subsurface differences between active and quiet regions: the stronger the surface magnetic field, the smaller the magnetic effects in the deeper layers, and vice versa \cite{Lin08}. Moreover, the magnetic effects in the deeper layers are the strongest in the quiet regions, consistent with the fact that these are basically regions with weakest magnetic fields at the surface.
\begin{figure}[htb!]
 \centerline{
   \includegraphics[width=0.535\textwidth, angle =90]{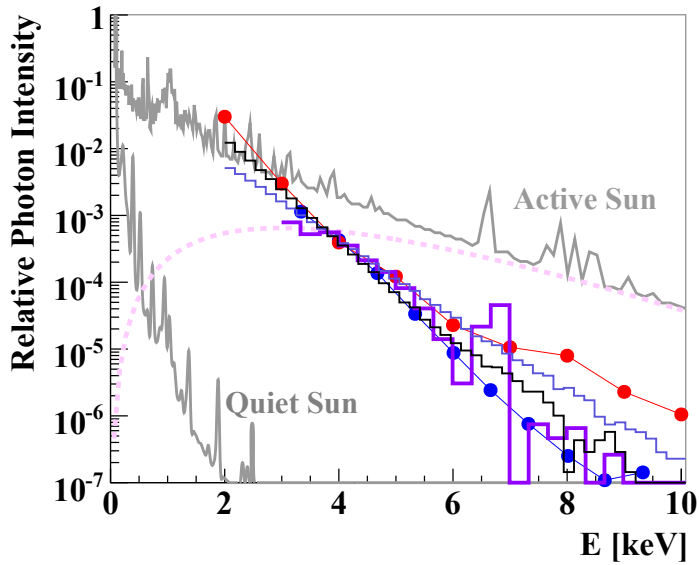}}
\caption{Solar X-ray spectra: Quiet and Active Sun (reconstructed)\cite{Zioutas:2009bw}. The other directly measured  spectra (RESIK, RHESSI) refer to a flare (red)\cite{Syl05}, non-flaring Sun (blue)\cite{mcTiernan} and pre-flaring periods (purple) after having subtracted the main X-ray flare component \cite{Bat09}, whose distribution coincides with that of the `Active Sun' above 3\,keV (not shown). Two thin lines, black and blue, show the expected shape of reconverted outstreaming solar axions at a depth of 400\,km and 350\,km below the solar surface, respectively. The dashed line shows the initial X-ray spectrum from converted solar axions (for a recent directly measured quiet Sun X-ray spectrum, see \cite{sylPage}). {\it Note: the distributions are not to scale, emphasis is given to their shape.}}
\label{figure}
\end{figure}

Figure \ref{figure} gives the analog spectra from the quiet and flaring Sun as well as from non-flaring active regions, including preflare periods. All spectra show actually a strikingly similar linear distribution in the log-lin plot, as expected from Geant4 simulation \cite{Zioutas:2009bw}. Especially, the recently directly measured quiet Sun spectrum above 1\,keV with the SphinX mission \cite{sylPage} confirms a perfect linear shape. The figure also gives the directly measured spectral shape of the emitted X-rays from the non-flaring Sun \cite{mcTiernan}, a solar flare \cite{Syl05} and pre-flaring periods \cite{Bat09}. It is interesting to point out here that nothing requires a solar emission to be always from a high-temperature plasma, though it is usually offered as the simplest explanation \cite{private}, even more so if its energy source is unknown (e.g. the solar corona heating problem). Only their power law exponent is different from case to case (Fig. \ref{figure}), with the quiet and flaring Sun being the two extreme cases, which is reasonable following the axion scenario \cite{Zioutas:2009bw}. Furthermore, it is widely accepted that flares are magnetic in origin, although their trigger remains elusive \cite{arxiv}. Interestingly, the peak of a flare X-ray intensity vs. B$_{max}$ indicates a B$^2$ dependence \cite{proc}. In particular, the non-flaring active region AR7978 provided an excellent L$_{x}$ vs. B$^2$ dependence, as it is expected from the Primakoff effect \cite{ar}; to have such a dependence for the soft X-rays, where the energy distribution from converted solar axions gets reduced (dashed line in Fig. \ref{figure}), an energy degradation is required. The suggested multiple Compton scatterings  allows this to happen \cite{Zioutas:2009bw}.

\section{Conclusion}

The considered solar analog spectra fit an axion scenario starting a few 100\,km below the solar surface, where outstreaming solar axions from the hot core, with a rest mass around 17\,meV/c$^2$, can coherently convert to X-rays. The exact depth varies from case to case, depending on the actual solar conditions like density and magnetic field configuration, which, in addition, change dynamically continuously. The quiet/non-flaring solar X-ray brightness can be qualitatively (spectrum shape) and quantitatively (yield) reconstructed without the need to invent new physics beyond that of the standard QCD axions. However, to explain quantitatively also the active or even flaring Sun within the same scenario, one is forced to assume (much) stronger magnetic fields and/or larger conversion lengths due to an occasionally occurring `fine tuning' between local density and axion(-like) rest mass; that this may happen occasionally is not far-fetched, since the more rare these `explosive' solar events are,the more powerful they appear.


\begin{footnotesize}

\end{footnotesize}


\end{document}